\documentclass[12pt]{article}

\begin{document}

\title{Thermostring Quantization. An Interpretation of Strings as Particles at
Finite Temperatures\thanks{%
Published in ''Z.Zakir (2003) {\it Structure of Space-Time and Matter. }%
CTPA, Tashkent.''}}
\author{Zahid Zakir \\
CTPA, P.O.Box 4412, Tashkent, 700000 Uzbekistan,\\
zahid@in.edu.uz}
\date{September 23, 1998;\\
Revised October 12, 2003.}
\maketitle

\begin{abstract}
An instantaneous temperature path of a point particle in the space -
temperature manifold can be represented as a string of length L=1/kT
(thermostring). The thermostring swepts a surface in the
space-time-temperature manyfold at its temporal evolution. The thermostring
is closed, its points can be rearranged and the charge is distributed along
the length. Some predictions of this method for statistical mechanics and
string theories are discussed.
\end{abstract}

\section{Introduction}

The well known analogy between quantum mechanics and quantum statistical
mechanics of particles under the replacements $\Delta t\rightarrow -i\Delta
\beta ,$ $\Delta \beta \rightarrow i\Delta t$ has been commonly used as a
formal procedure at the consideration one of $t-$ , or $\beta -$evolutions 
\cite{1}. Here $\Delta t-$ is a temporal interval, $\Delta \beta -$ is
(inverse) temperature evolution parameter's interval (''cold'') $0\leq
\Delta \beta \leq 1/kT$, $T-$ is temperature, $k-$ is the Boltzmann
constant. However, if we consider $\beta -$evolution of particles in {\it %
combination} with $t-$evolution, this analogy leads to some $(t-\beta )-$ 
{\it symmetry.} Here the inverse temperature parameter $\beta $ can be
treated as a new geometrical degree of freedom in addition to the spatial
and temporal ones. So, as a result, we have some space-time-temperature
manifold with $D=d+2$ dimensions, where $d$ is space dimensionality. We
shall introduce in this manifold {\it the} {\it thermostring representation
of density matrix }or{\it \ the} {\it thermostring quantization} (see also 
\cite{Za},\cite{3}). For non-relativistic particles the thermostring
quantization leads to some reformulation of the statistical mechanics. The
thermostring quantization of relativistic particles at some invariant
temperature $T_{0}$, corresponding to some invariant length, for example, to
the Planck length and charge radii of particles, leads to the closed string
formalism. Here the invariance of the temperature $T_{0}$ is very important
and this fact allows us to treat the temperature degree of freedom as one of
geometrical dimensions of the physical manifold for the given system. The
thermostring quantization makes it possible a new physical interpretation of
superstrings IIA and heterotic strings in terms of point particles.

\subsection{Thermomechanics of particles}

The density matrix $\rho $ for {\it mixed} states of a non-relativistic
particle at the finite temperature $T$ can be represented as: 
\begin{eqnarray}
\rho ({\bf r},{\bf r}_{0};\Delta \beta ) &=&\sum\limits_{i}\exp
(-E_{i}\Delta \beta )\times \psi _{i}({\bf r})\psi _{i}^{\ast }({\bf r}_{0})=
\nonumber \\
&=&\sum\limits_{i}\psi _{i}({\bf r},\beta )\psi _{i}^{\ast }({\bf r}%
_{0},\beta _{0})=\rho ({\bf r},\beta ;{\bf r}_{0},\beta _{0}),
\end{eqnarray}
where $\Delta \beta =\beta -\beta _{0}=1/kT,$ $E_{i}$- is the energy of
particle, ${\bf r}$ is a spatial radius-vector. Here the wave functions $%
\psi _{i}({\bf r},\beta )$ describe {\it pure} states of the particle in a $(%
{\bf r},\beta )$-manifold with the Hamiltonian $H$: 
\begin{eqnarray}
\psi _{i}({\bf r},\beta ) &=&\exp (-H\beta )\psi _{i}({\bf r}),  \nonumber \\
\psi _{i}^{\ast }({\bf r},\beta ) &=&\psi _{i}^{\ast }({\bf r})\exp (H\beta
),
\end{eqnarray}
and normalized as:

\begin{equation}
\int \psi _{i}^{\ast }({\bf r},\beta )\psi _{j}({\bf r},\beta )d{\bf r}%
=\delta _{ij}.  \label{Norm}
\end{equation}

In $({\bf r},\beta )-$manifold, the density matrix $\rho $ plays the role of
a transition amplitude for the $\beta -$evolution of the wave functions $%
\psi _{i}({\bf r},\beta )$:

\begin{equation}
\psi _{i}({\bf r},\beta )=\int d{\bf r}_{0}\rho ({\bf r},\beta ;{\bf r}%
_{0},\beta _{0})\psi _{i}({\bf r}_{0},\beta _{0}),
\end{equation}
with the property:

\begin{equation}
\rho ({\bf r},\beta ;{\bf r}_{0},\beta _{0})=\int d{\bf r}_{1}\rho ({\bf r}%
,\beta ;{\bf r}_{1},\beta _{1})\rho ({\bf r}_{1},\beta _{1};{\bf r}%
_{0},\beta _{0}).
\end{equation}
Its path integral representation is \cite{1}:

\begin{equation}
\rho ({\bf r}_{n},\beta _{n};{\bf r}_{0},\beta _{0})={\int\limits_{{\bf r}%
(\beta _{0})}^{{\bf r}(\beta _{n})}}D{\bf r}(\beta )\exp \left[ -\frac{m}{2}{%
\int\limits_{\beta _{0}}^{\beta _{n}}}({\bf r}^{\prime })^{2}d\beta \right] ,
\end{equation}
where ${\bf r}^{\prime }=d{\bf r/}d\beta $. The partition function $Z(\Delta
\beta )$ is defined as: 
\begin{equation}
Z(\Delta \beta )=\int d{\bf r}\rho ({\bf r},\beta ;{\bf r}_{0},\beta
_{0})\mid _{{\bf r}(\beta )={\bf r}(\beta _{0})}.
\end{equation}

In this representation of the density matrix:

a) The states of particles are described in the space-temperature manifold $(%
{\bf r},\beta )$, i.e. the thermal degree of freedom is treated as a
geometrical dimension of the manifold in addition to the spatial and
temporal ones. Distances along this axis are proportional to the Planck
constant $\hbar $ and inverse proportional to the temperature $T$;

b) Mixed states for the statistical ensemble of point particles are
described by the density matrix $\rho ({\bf r},{\bf r}_{0};\beta )$, which
after the factorization $\Delta \beta =\beta -\beta _{0}$ may be represented
as the transition amplitude $\rho ({\bf r},\beta ;{\bf r}_{0},\beta _{0})$
for the ''pure'' states $\psi _{i}({\bf r},\beta )$ in the $({\bf r},\beta )$%
- manifold;

c) The temperature parameter of evolution $\beta $ is unlimited $-\infty
\leq \beta \leq \infty ,$ but the intervals $\Delta \beta $ are restricted
as: $0\leq \Delta \beta \leq 1/kT$, and therefore, the temperature paths of
the particles have finite (mean) lengths;

d) Only closed temperature paths with ${\bf r}(\beta _{n})={\bf r}(\beta
_{0})$ contribute to the partition function, and we have for their
coordinates the periodicity condition ${\bf r}(\beta )={\bf r}(\beta +1/kT)$.

In Feynman's path integral representation of the density matrix \cite{1}
there exists the asymmetry between the temporal and thermal degrees of
freedom, because of the temperature evolution is described by the path
integral, whereas the time evolution is described by the differential
equations. Further we shall consider another method for the summation of
probabilities for the temperature paths, where both thermal and temporal
evolutions are described by the path integrals.

\subsection{The temperature paths as thermostrings}

The density matrix as the $\beta $-evolution transition amplitude can be
represented through infinity set of intermediate states ($n\rightarrow
\infty $):

\begin{eqnarray}
\rho ({\bf r}_{n},\beta _{n};{\bf r}_{0},\beta _{0}) &=&\int d{\bf r}(\beta
_{1})...d{\bf r}(\beta _{n-1})\times  \nonumber \\
&&\times \rho ({\bf r}_{n},\beta _{n};{\bf r}_{n-1},\beta _{n-1})\times ... 
\nonumber \\
&&...\times \rho ({\bf r}_{2},\beta _{2};{\bf r}_{1},\beta _{1})\rho ({\bf r}%
_{1},\beta _{1};{\bf r}_{0},\beta _{0}),
\end{eqnarray}
which means that:

\begin{eqnarray}
\rho ({\bf r}_{n},\beta _{n};{\bf r}_{0},\beta _{0}) &=&\int d{\bf r}(\beta
_{1})...d{\bf r}(\beta _{n-1})\times   \nonumber \\
&&\{\psi _{i}({\bf r}_{n},\beta _{n})\psi _{i}^{\ast }({\bf r}_{n-1},\beta
_{n-1})\psi _{j}({\bf r}_{n-1},\beta _{n-1})\psi _{j}^{\ast }({\bf r}%
_{n-2},\beta _{n-2})\times ...  \nonumber \\
&&...\times \psi _{k}({\bf r}_{2},\beta _{2})\psi _{k}^{\ast }({\bf r}%
_{1},\beta _{1})\psi _{i}({\bf r}_{1},\beta _{1})\psi _{i}^{\ast }({\bf r}%
_{0},\beta _{0})\}.
\end{eqnarray}
Here the summation over the repeated indices $i,j,k,...$ is supposed. Due to
the normalization conditions, there contribute only the intermediate states
with the indices equal to the initial and final states indices: $i=j=k=...$

Therefore, we can transfer all $\psi _{i}({\bf r}_{p},\beta _{p})$ to left
hand side and all $\psi _{i}^{\ast }({\bf r}_{p},\beta _{p})$ to right hand
side: 
\begin{eqnarray}
\rho ({\bf r}_{n},\beta _{n};{\bf r}_{0},\beta _{0}) &=&\int d{\bf r}(\beta
_{1})...d{\bf r}(\beta _{n-1})\times   \nonumber \\
&&\{\psi _{i}({\bf r}_{n},\beta _{n})\psi _{i}({\bf r}_{n-1},\beta
_{n-1})...\psi _{i}({\bf r}_{1},\beta _{1})\}\times   \nonumber \\
&&\times \{\psi _{i}^{\ast }({\bf r}_{n-1},\beta _{n-1})...\psi _{i}^{\ast }(%
{\bf r}_{1},\beta _{1})\psi _{i}^{\ast }({\bf r}_{0},\beta _{0})\},
\end{eqnarray}
and then rewrite this expression across the wave functionals $\Psi _{i}$and $%
\Psi _{i}^{\ast }$ as: 
\begin{equation}
\rho ({\bf r}_{n},\beta _{n};{\bf r}_{0},\beta _{0})={\int\limits_{{\bf r}%
(\beta _{0})}^{{\bf r}(\beta _{n})}}D{\bf r}(\beta )\Psi _{i}[{\bf r}(\beta
);\beta _{n},\beta _{0}]\Psi _{i}^{\ast }[{\bf r}(\beta );\beta _{n},\beta
_{0}].
\end{equation}
Here $D{\bf r}(\beta )$ is the path integration measure, the wave
functionals $\Psi _{i}$ and $\Psi _{i}^{\ast }$ describe a state of the
temperature path and are composed as the product of infinity number ($%
n=(\beta _{n}-\beta _{0})/\varepsilon \rightarrow \infty $) intermediate
state wave functions $\psi _{i}$ of the same energy $E_{i}$ as: 
\begin{eqnarray}
\Psi _{i}[{\bf r}(\beta );\beta ,\beta _{0}] &=&\lim\limits_{n\rightarrow
\infty }\prod\limits_{k=0}^{n}\psi _{i}({\bf r}_{k},\beta _{k}), \\
\Psi _{i}^{\ast }[{\bf r}(\beta );\beta ,\beta _{0}]
&=&\lim\limits_{n\rightarrow \infty }\prod\limits_{k=0}^{n}\psi _{i}^{\ast }(%
{\bf r}_{k},\beta _{k}).
\end{eqnarray}

We see that the temperature path of the point particle ($\beta -$world line)
can be described as some one dimensional physical object in
space-time-temperature manifold with the wave functional $\Psi _{i}$, and
further we shall call this object as {\it the thermostring.}

In general case the wave functional $\Psi _{i}$ must be symmetrized under
the permutations of points of the temperature path (thermostring). These
permutations depend on a type of statistics of particles in the Gibbs
ensemble, and they determine the type of statistics of the thermostring
(bosonic or fermionic). In the ordinary string theory the such permutations
impossible since the ordinary strings are introduced as continuous one
dimensional objects in the physical space.

\subsection{The temporal evolution of thermostrings}

During a time interval the temperature path (thermostring) swepts a surface
in the space-time-temperature manifold. An action function for the particle
at the finite temperature and the temporal evolution of the corresponding
density matrix can be described by the summation over all the such surfaces.
This circumstance allows us to introduce the thermostring representation of
the quantum statistical mechanics of particles or {\it the thermostring
quantization} of the particle at finite temperatures.

The standard time dependence of the wave functions:

\begin{equation}
\psi _{i}({\bf r},\beta ,t)=\exp (-iHt)\psi _{i}({\bf r},\beta ),
\end{equation}
allows us to obtain a time dependence of the wave functionals $\Psi _{i}.$

In $({\bf r},\beta )-$ manifold we have $(d+1)$-dimensional spacelike
coordinates ${\bf q}$ with components $({\bf r},\beta )$. Temporal
derivatives of these vectors are space-temperature velocities of the
particles, and they can be separated into longitudinal and transverse to the
temperature path components: 
\begin{eqnarray}
\frac{\partial {\bf q}}{\partial t} &\equiv &{\bf v=v}_{\perp }+{\bf v}%
_{\parallel }, \\
{\bf v}_{\perp } &=&{\bf v}-{\bf k}({\bf q}^{\prime }{\bf v}),  \nonumber
\end{eqnarray}
where ${\bf k}={\bf q}^{\prime }/({\bf q}^{\prime })^{2}$ and ${\bf q}%
^{\prime }=\partial {\bf q}/\partial \beta .$

The longitudinal components of the velocity ${\bf v}_{\parallel }$ also have
two parts. The first one leads to the collective motion of the thermostring
as a whole object with synchronous displacements of all points of the
thermostring, and in the case of a single thermostring these displacements
can be disregarded (as zero modes). The second part of the longitudinal
velocity leads to permutations of neighbor points of the thermostring. These
permutations do not contribute to the thermostring energy because of
indistinguishability of its points. In the string theory the exclusion of
the ${\bf v}_{\parallel }$ from the Lagrangian is one of difficulties of the
theory \cite{4}, whereas in the case of thermostrings this is a natural and
necessary procedure.

So, we have following time dependence formula for the wave functional:

\begin{equation}
\Psi _{i}[{\bf q}(\beta ,t),\beta _{n},\beta _{0};t]=\exp \left( -\frac{%
i\Delta t}{\Delta \beta }{\int\limits_{\beta _{0}}^{\beta _{n}}}d\beta H(%
{\bf v}_{\perp }^{2})\right) \Psi _{i}[{\bf q}(\beta );\beta _{n},\beta
_{0};t_{0}].
\end{equation}

This expression can be represented in the form of a surface integral as:

\begin{equation}
\Psi _{i}[{\bf q}(\beta ,t);\Delta \beta ;t_{n}]=P[{\bf q}(\beta ,t);\Delta
t;\Delta \beta ]\Psi _{i}[{\bf q}(\beta ,t_{0});\Delta \beta ;t_{0}],
\end{equation}
where: 
\begin{eqnarray}
P[{\bf q}(\beta ,t);\Delta t;\Delta \beta ] &=&\int D{\bf q}(\beta ,t)\exp
\left( \frac{i}{\Delta \beta }\int\limits_{\beta _{0},t_{0}}^{\beta
_{n},t_{n}}d\beta dtL[{\bf v}_{\perp }(\beta ,t),{\bf q}(\beta ,t)]\right) =
\nonumber \\
&=&\int D{\bf q}(\beta ,t)\exp \left\{ iS[{\bf q}(\beta ,t);\Delta t;\Delta
\beta ]\right\}
\end{eqnarray}

Here $P[{\bf q}]$ is a propagator and $S[{\bf q}]$ is the action function
for the thermostring:

\begin{eqnarray}
S[{\bf q}] &=&\frac{1}{\Delta \beta }\int d\beta dtL[{\bf v}_{\perp }(\beta
,t),{\bf q}(\beta ,t)]=  \nonumber \\
&=&\frac{m}{2\Delta \beta }\int d\beta dt\times {\bf v}_{\perp }^{2}.
\end{eqnarray}

For the Gibbs ensemble of free relativistic particles we have following
action function for the corresponding relativistic thermostrings: 
\begin{equation}
S[x]=-\frac{m}{\Delta \beta }\int d\beta dt\sqrt{1-{\bf v}_{\perp }^{2}},
\end{equation}
where $(d+2)$-vector $x^{\mu }$ have components $x^{\mu }({\bf q,}t)=x^{\mu
}({\bf r,}\beta ,t{\bf ).}$

As it can be shown \cite{4}, after the introduction of the world sheet
parameters $\tau ,\sigma $ as $t(\tau ,\sigma )$, $\beta (\tau ,\sigma ),$
and the substitutions: 
\begin{eqnarray}
d\beta  &=&d\sigma \sqrt{x^{\prime 2}}, \\
dtd\sigma  &=&\frac{\partial (t,\sigma )}{\partial (\tau ,\sigma )}d\tau
d\sigma =\stackrel{.}{t}d\tau d\sigma ,
\end{eqnarray}
where $\stackrel{.}{x}^{\mu }=\partial x^{\mu }/\partial \tau ,x^{\prime
}=\partial x/\partial \sigma ,$ $\partial {\bf q/}\partial t=\stackrel{.}{%
{\bf q}}/\stackrel{.}{t},$ and $\partial {\bf q/}\partial \sigma ={\bf q}%
^{\prime }-\stackrel{.}{{\bf q}}t^{\prime }/\stackrel{.}{t}$, the action
function transforms into the Nambu-Goto action for the relativistic
thermostring:

\begin{equation}
S[x]=-\gamma {\int }d\sigma d\tau \sqrt{(\stackrel{.}{x}x^{\prime })^{2}-%
\stackrel{.}{x}^{2}x^{\prime 2}},
\end{equation}
where $\gamma =m/\Delta \beta =mkT$. We see that in the thermostring
representation of quantum statistical mechanics of particles there exist the
reparametrizational symmetry $\sigma ^{\prime }=f(\sigma ,\tau ),$ $\tau
^{\prime }=\varphi (\sigma ,\tau )$ as in the string theories.

Another form of the action function for the relativistic particle leads to
the second form of the action function for the thermostrings \cite{2}:

\begin{equation}
S[x,e]=-\frac{1}{2\Delta \beta }\int d\beta dt(e^{-1}{\bf v}_{\perp }^{2}-em)
\end{equation}

This action is fully relativistically invariant if $\Delta \beta $ and the
limits of the $\beta $-integration are invariants. In ordinary temperatures
it is impossible, but if we take the relativistically invariant (Planck,
charge radii, etc.) temperature $T_{0}$ as a limiting temperature for the
thermostrings with $\Delta \beta _{0}=1/kT_{0}$, then we have the invariant
action function. In terms of the world-sheet coordinates $x(\sigma ,\tau )$
and metrics $g(\sigma ,\tau )$ this action leads to the Polyakov surface
integral \cite{5} for the thermostring amplitude:

\begin{equation}
P[x,g]=\int Dx(\sigma ,\tau )Dg(\sigma ,\tau )\exp (iS[x,g])
\end{equation}

Finally, we conclude that the statistical mechanics of the relativistic
particle at finite temperatures in the thermostring representation is
identical with the closed string theory formalism.

\subsection{Statistical mechanics in the thermostring representation}

The thermostring representation may be useful for the solving of equilibrium
and non-equilibrium statistical mechanics problems by means of new methods
of the string theory formalism. The surface integral representation of the
Green functions of statistical mechanics and the geometric formulation can
lead to some nontrivial consequences.

As an example can be considered the problem of doubling of operators in the
thermofield dynamics, which developed by H.Umezawa et al. In the
thermostring representation this doubling can be explained as the
consequence of the existence of left and right moving modes for the closed
thermostrings. The thermal algebra for these doubled states can be
represented as the Virasoro algebra for the thermostring modes.

Another nontrivial consequence of the thermostring quantization may be the
appearance of the Liouville modes and tachyons as additional fields or
collective excitations for some statistical systems. The Liouville modes of
the thermostrings in statistical mechanics appear because of the
dimensionality of the manifold is less than the critical dimensionality of
the manifold (10 or 26). Tachyons also may appear in the thermostring
quantization of some statistical systems due to the absence of fermions or
supersymmetry \cite{2}.

\subsection{Thermostring interpretation of closed strings}

In the standard string theory the strings have been introduced as new
fundamental one dimensional objects in the physical space. The introduction
into physics of the such non-local structures leads to conceptual
difficulties concerning measurability of points of the strings, their
intrinsic dynamics, relativistic causality, etc. There is a disconnection
between the pointness of the space-time manifold with metrics and non-local
nature of the strings, some modes of which generate gravitons. Strings and
other their generalizations, such as branes, are {\it alien} to the geometry
of space-time and the reduction of the strings and branes to space-time
properties is nontrivial problem.

The thermostring quantization does not introduce new objects into physics,
the dimensionality of the space is lower to one (with respect to the string
theories), it exactly coincides with the string theory formalism, and it can
be used directly for the description of the physical systems without any
hypothesis. This is only a new method of quantization at finite temperatures
of the ordinary objects - point particles and local fields, where the
thermostring appears as some non-local structure in the configurational {\it %
space-- time- temperature} manifold. The physical reason for non-locality is
the Gibbs ensemble averaging and the thermostring length is nothing but as
the instantaneous length of the space-temperature world-line of the point
particle. So, in the thermostring representation we introduce into physics
only a new (temperature) dimension in addition to the space and time, and we
represent the physical manifold as the space-time-temperature manifold with
more general properties than in space-time.

The superstrings \cite{2} can be interpreted as thermostrings due to the
identity of the formalisms by the preserving all achievements of the string
theory. At the same time, the new treatment allows one to exclude some
conceptual difficulties associated with the introduction into physics of
non-local fundamental objects with immeasurable intrinsic structure.

Thermostring interpretation of the superstrings leads to the following
consequences:

a) Strings are not fundamental physical objects in the physical space, they
are point particles moving in the fluctuating background with some invariant
temperature. They are described by the string formalism because of the need
for Gibbs ensemble averaging;

b) One of dimensions of the string theory manifold is the thermal dimension,
and, therefore, the dimensionality of the space is equal to 8 (or 24),
which, together with temporal and thermal degrees of freedom, form the
critical dimensionality;

c) Only the closed strings appear at the initial and final states of the
string theory amplitudes as observable physical states;

d) The charge of the particle must be distributed along the thermostring;

e) The interactions of the strings with point particles become more clearly
understandable;

f) The invariant temperature $T_{0}$ can be considered at large distances
(and low temperatures) as very large quantity $T_{0}\rightarrow \infty $,
and the thermostrings with $\Delta \beta \rightarrow 0$ become like to a
point particle.

Among superstring theories only theories of closed strings satisfy to these
conditions, and we may conclude that only the superstrings IIA (in the case
of neutral particles) and the heterotic strings (in the case of charged
particles) can be interpreted as thermostrings.

The thermostring interactions may be described simultaneously in the
particle, statistical ensemble and string languages. The factorization of
one a statistical ensemble into two ensembles, or the merging of two
ensembles into one are physically clearer and simpler procedures than the
cutting or gluing of rigid strings in the string theory.

We can perform the thermostring quantization of {\it physical strings} as
one dimensional objects in physical space and we shall obtain as a result
the theory of {\it thermo-membranes}. So, if we provide the thermostring
quantization of the physical D-branes, we shall obtain the theory of
(D+1)-thermobranes, i.e. the dimensionality of initial objects of the theory
increases to one. In the treatment of the D-branes the thermostring
representation can be combined with M-Theory methods if we interpret one of
it's 11 dimensions as temperature degree of freedom.

Finally, we conclude that the thermostring is not only simplest and natural,
but necessary modification of the local theories at small distances and high
temperatures.

\end{document}